\documentclass{article}
\usepackage{amsmath}
\usepackage{graphicx}
\usepackage{amsfonts}
\usepackage{amssymb}

\def\beq{\begin{equation}}
\def\eeq{\end{equation}}
\def\bea{\begin{eqnarray}}
\def\eea{\end{eqnarray}}

\def\q {q}

\begin{document}

\begin{center}

\Large{CENTRAL NUCLEON-NUCLEON POTENTIAL}

\Large{AND CHIRAL SCALAR FORM FACTOR}

\end{center}

\vspace{5mm}

\baselineskip=5mm

\begin{center}

M.R. ROBILOTTA

\it\small{Nuclear Theory and Elementary Particle Phenomenology Group, Instituto de F\'{\i}sica,\\
Universidade de S\~{a}o Paulo, C.P. 66318, 05315-970, S\~ao Paulo, SP, Brazil}

\end{center}

\vspace{5mm}

\begin{abstract}

The central two-pion exchange $NN$ potential at large distances is studied in the framework of relativistic chiral symmetry and related directly to the nucleon scalar form factor, which describes the mass density of its pion cloud.
This relationship is well supported by phenomenology and allows the dependence of the asymptotic potential on the nucleon 
mass to be assessed.
Results in the heavy baryon limit are about 25\% larger than those corresponding to the empirical nucleon mass 
in the region of physical interest.
This indicates that it is very important to keep this mass finite in precise descriptions of the $NN$ system
and supports the efficacy of the relativistic chiral framework.
One also estimates the contribution of subleading effects and presents a simple discussions of the role of the quark condensate 
in this problem.

\end{abstract}

\vspace{5mm}

\section{Introduction}

QCD is nowadays the main theoretical framework for understanding hadronic processes, but its non-Abelian character makes low energy calculations unfeasible.
The usual strategy for overcoming this difficulty consists in working with effective theories, constructed in such a way as to include,
as much as possible, the main features of QCD.
As in Nuclear Physics most processes involve only the quarks $u$ and $d$, one requires these theories to be Poincar\'e invariant 
and to possess approximate $SU(2)\times SU(2)$ chiral symmetry.
The breaking of the latter, due to the quark masses, is parametrized by the small pion mass $(\mu)$ at the effective level.

In the last twenty years, effective theories incorporating chiral symmetry have been successfully applied to hadronic interactions, 
including or not electro-weak probes.
As far as the hadronic sector is concerned it is useful to distinguish two classes of processes, involving only mesons  or both mesons and baryons.
In the case of purely pionic systems, effective Lagrangians are treated relativistically and yield well defined power counting 
procedures, in both pion mass and momenta \cite{W79, GL}.
This gives meaning to the idea of  chiral perturbation theory.
When nucleons are present, things become more complicated \cite{GSS} and calculations were performed in the
simplified framework of the so called heavy baryon chiral perturbation theory (HBChPT), in which nucleons
are treated non-relativistically \cite{WNN, JM}.
Only recently a well defined power counting scheme was proposed for the relativistic approach \cite{BL}.

In the case of two-nucleon systems, interactions involve processes with a marked spatial hierarchy, whose dynamical complexity increases rapidly when one moves inward.
In the best semi-phenomenological potentials existing at present, that can reproduce low-energy observables accurately, 
the interaction is determined by the undisputed one pion exchange, supplemented by a theoretical 
two-pion exchange potential (TPEP) and parametrized at short distances \cite{P,B}.

The TPEP is the locus of chiral symmetry in this problem, since it is closely related to the pion-nucleon ($\pi$N) amplitude \cite{P}.
After the works of Weinberg in the early nineties restating the role of chiral symmetry in nuclear interactions \cite{WNN}, the interest in applications was renewed and several authors have tackled the problem of constructing the TPEP. 
Initially, non-linear Lagrangians containing only pions and nucleons were employed \cite{ChNN}. 
However, these Lagrangians do not suffice to describe experimental $\pi N$ data \cite{H83} and the corresponding potential misses even the scalar-isoscalar medium range NN attraction.
In a later stage, other degrees of freedom were introduced, coherence with $\pi N$ information 
was restored and descriptions compatible with asymptotic NN scattering data were produced \cite{ORK, RR, KBW, KGW,BRR}.

It is possible to distinguish two partially overlapping theoretical frameworks in these chiral calculations.
One of them, adopted in refs.\cite{ORK} and \cite{KBW}, is based on HBChPT.
In this theory, a rule exists for counting powers of  the typical three-momenta exchanged between nucleons, assumed to be of the order of the pion mass.
This allows the construction of non-relativistic effective Lagrangians, that include unknown couterterms and are used to derive amplitudes in which contributions from loops and these counterterms add up coherently.
A rather puzzling aspect of these counting schemes, in the case of NN interactions, is that they predict a leading contribution in the $1/m$ expansion with the following spin-isospin structure \cite{KBW}:
$V_{leading}= V_S \;\mbox{\boldmath $\sigma$}^{(1)}\!\cdot\!\mbox{\boldmath $\sigma$}^{(2)} + V_T \; S_{12} + W_C \; \mbox{\boldmath $\tau$}^{(1)}\!\cdot\!\mbox{\boldmath $\tau$}^{(2)}$,
where $\mbox{\boldmath $\tau$}$, $\mbox{\boldmath $\sigma$}$ and $S_{12}$ are the usual isospin, spin and tensor operators.
This means that, in HBChPT, the scalar-isoscalar central potential, which is by far the most important phenomenological component of the TPEP, corresponds to just a subleading effect.

The other approach to the TPEP is typically relativistic \cite{RR} and emphasis is given to the tail of the interaction, which is determined by the well known $\pi$N amplitude.
No attention is paid to counterterms in the effective Lagrangian, that correspond to zero-range interactions and represent processes such as multimeson or quark exchanges.
Tail and conterterm contributions do not overlap in space and hence can be treated separately in NN potentials.

For systems containing just a single nucleon, relativistic and HBChPT calculations were compared and found to be
consistent, provided the dimensional regularization scale is set equal to the nucleon mass  \cite{BKKM}.
In this work we repeat this kind of comparison, for the leading contribution to the two-pion exchange $NN$ potential, 
and show that relativistic and HBChPT predictions fail to agree by 30\%.
The reason for this important discrepancy may be traced back to the fact that the intermediate $\pi N$ amplitude cannot be represented by a series around the point that determines the potential at large distances, as shown by Becher and Leutwyler \cite{BL}.

\section{TPEP}

The construction of the TPEP was discussed in detail in ref. \cite{RR} and here we just sketch the main steps.
The potential is based on the on-shell NN scattering amplitude containing two intermediate pions, from which 
one subtracts the iterated OPEP, in order to avoid double counting.
The isoscalar component, represented by ${\cal T}^S$, is given by 
\begin{equation}
{\cal T}^S = -\;\frac{i}{2!}\int \frac{d^4Q}{(2\pi)^4}\;\frac{ 3\; [T^+]^{(1)}\;[T^+]^{(2)}}  {[k^2-\mu^2]\;[k^{\prime\,2}-\mu^2]}-(OPEP)^2 \;,
\label{1}
\end{equation}
where $T^+$ is the isospin symmetric $\pi N$ scattering amplitude for pions with momenta $k$ and $k'$, $Q=(k'+k)/2$, 
and the factor $1/2!$ accounts for the symmetry under the exchange of the intermediate pions \cite{RR,MPR}.
In the sequence we represent initial and final nucleon momenta by $p$ and $p'$, their mass by $m$ and also use the variables $\q=(k'-k)$, $t=\q^2$, $V=(p'+p)/2m$ and $\nu=V \!\cdot\! Q$.

In general, the amplitude $T^+$ can depend on four independent variables, $\nu$, $t$, $k^2$ and $k^{\prime\,2}$.
As the exchanged pions are off shell, one should, in principle, keep $k^2$ and $k^{\prime\,2}$ unconstrained everywhere.
However, when these factors appear in the numerator of amplitudes, we may write $k^2=\mu^2+(k^2-\mu^2)$  
or $k^{\prime\,2}=\mu^2+(k^{\prime\,2}-\mu^2)$  and use the terms within parentheses to cancel pion propagators in eq.(\ref{1}). 
This kind of cancellation is associated with short range terms, which do not contribute to the asymptotic potential.
Therefore, whenever possible, we neglect them and replace $k^2$ and $k^{\prime\,2}$ by $\mu^2$ in the numerators of our expressions.
Concerning the variables $\nu$ and $t$, the conditions of integration in eq.(\ref{1}) are such that the main contributions 
come from the unphysical region $\nu\approx 0$, $t\ge 4\mu^2$ \cite{P}. 
In particular, the vicinity of the point $t=4\mu^2$ determines the potential at very large distances.

In order to understand the structure of the TPEP, it is convenient to isolate in $T^+$ a term $T_N^+$, involving only chiral pion-nucleon interactions at tree level and written as
\begin{equation}
T_N^+ =  \frac{g^2}{m}\;\bar{u} \left\{ 1 -
\left[ \frac{m}{(p+k)^2-m^2}-\; \frac{m}{(p-k')^2-m^2}\right] \not\!Q \right\} u  \;, 
\label{2}
\end{equation}
where $g$ is the $\pi N$ coupling constant.
The remainder is represented by $T_R^+$ and encompasses other degrees of freedom, either in the form of low energy constants 
or by means of specific models, that may include the $\Delta$ and other states.
One writes symbolically $[T^+]=[T_N^+] + [T_R^+]$ for each nucleon and the potential is then proportional to 
$[T^+]^{(1)}[T^+]^{(2)}=[T_N^+]^{(1)}[T_N^+]^{(2)}
+\{ [T_N^+]^{(1)}[T_R^+]^{(2)} +[T_R^+]^{(1)}[T_N^+]^{(2)}\} +[T_R^+]^{(1)}[T_R^+]^{(2)}$.
This product involves box, crossed box, triangle and bubble diagrams, that should be taken into account in a complete calculation.
As pointed out by Gross \cite{Gr} long ago, the $[T_N^+]^{(1)}[T_N^+]^{(2)}$ contribution is very small, due to 
chiral cancellations.
The detailed numerical study of this structure performed in ref.\cite{RR} has confirmed this conclusion and shown that the 
crossed tem, within curly brackets, is largely dominant.
It is due to a $T_R^+$ which may be expressed as 
\begin{equation}
T_R^+ =  \left( a_{00}^+ +t\;a_{01}^+ \right) \; \bar{u}\;u \;,
\label{3}
\end{equation}
where the $a_{0i}^+$ are subthreshold coefficients \cite{H83}.
In this multipolar expansion, $a_{01}^+$ represents the fact that a nucleon can be deformed when interacting with pions
and is associated with the so called axial polarizability parameter, given by $ \alpha_A = 2\; a_{01}^+$.
\begin{figure}[h]
\begin{center}
\includegraphics[width=10cm]{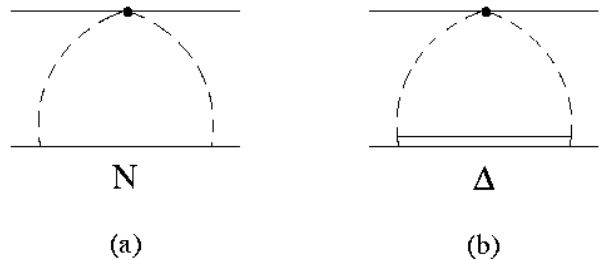}
\caption{Leading contributions to the two-pion exchange potential; the black dot indicates that one of the nucleons acts as 
scalar source, that disturbs the pion cloud of the other.}
\label{fig.1}
\end{center}
\end{figure}

This allows the asymptotic central potential to be written as 
\begin{equation}
{\cal T}_a^S(t) = \left[ \left( a_{00}^+ +t\;a_{01}^+ \right) \;\bar{u}\;u \right]^{(1)} \;\left[\frac{\sigma_N(t)}{\mu^2}\;\bar{u}\;u\right]^{(2)}+ \left(1 \leftrightarrow 2\right)\;,
\label{4}
\end{equation}
where $\sigma_N(t)$ is the nucleon contribution to the scalar form factor
\footnote {In ref.\cite{MPR} a similar expression was used, based on the parameter 
$\alpha_{00}^+  \equiv [\mu(a_{00}^+ +4\mu^2 \;a_{01}^+)]$, which tends to the present one for very long distances.}.
Quite generally, this form factor is defined in terms of the symmetry breaking Lagrangian ${\cal L}_{sb}$ as
$\langle p'|- {\cal L}_{sb}|p \rangle= \sigma(t)\; \bar{u}\;u $.
As discussed by Gasser, Sainio and \v{S}varc \cite{GSS}, $\sigma_N(t)$ is associated with the diagram of fig.\ref{fig.1}a and
given by 
\begin{eqnarray}
\sigma_N(t) = \frac{3}{2}\;\frac{g^2 \mu^2}{m}\; \frac{1}{(4\pi)^2}\; \left[ J_{c,c}(t) - J_{c,sN}^{(1)}(t) \right]  \;,
\label{5}
\end{eqnarray}
where the loop integrals  $ J(t)$ are written as 
\begin{eqnarray}
&& J_{c,c}(t) = \frac{(4\pi)^2}{i} 
\int\frac{d^4Q}{(2\pi)^4}\; 
\frac{1}{[(Q\!-\!\q/2)^2\!-\!\mu^2]\;[(Q\!+\!\q/2)^2\!-\!\mu^2]}\;,
\label{6}\\[2mm]
&& J_{c,sN}^{(1)}(t)  = \frac{(4\pi)^2}{i}\;\frac{1}{V^2} 
\nonumber\\[2mm]
&&\times \int\frac{d^4Q}{(2\pi)^4}\;
\frac{2mV\!\cdot\! Q}{[(Q\!-\!\q/2)^2\!-\!\mu^2]\;[(Q\!+\!\q/2)^2\!-\!\mu^2]\;
[Q^2\!+\!2mV\!\cdot\! Q\!-\!\q^2/4]}\;.
\label{7}
\end{eqnarray}

The integral given by eq.(\ref{6}) is divergent and has to be regularized. 
However, this procedure affects only the zero-range properties of the potential and is irrelevant for the present discussion,
which concentrates on asymptotic properties.
Therefore, in the sequence, we consider only differences between functions calculated at generic values of $t$ and at $t=0$,
denoted by a caret. 

The non-relativistic potential is obtained by going to the center of mass frame, where $t=-\mbox{\boldmath $\q$}^2$, and making $[ \bar{u} \; u ] \rightarrow1$ in eq.(\ref{4}).
One then has 
\begin{equation}
\hat{V}_a^S(m; \mbox{\boldmath $\q$}) = 
- 2\; \left( a_{00}^+ -\mbox{\boldmath $\q$}^2 \;a_{01}^+ \right)
\left\{ \left[ \frac{3 \mu}{32\pi}\left( \frac{g}{m}\right)^2 \right] 
\frac{m}{\pi \mu}  \left[ \hat{J}_{c,c}(m; \mbox{\boldmath $\q$}) - \hat{J}_{c,sN}^{(1)}(m; \mbox{\boldmath $\q$}) \right] \right\} \;,
\label{8}
\end{equation}
where we now indicate explicitly the dependence of the potential on the nucleon mass.
The minus sign in front of this expression was introduced in order to account for our convention of the relativistic T-matrix.
It is worth noting that this contribution was already included in the general formulation of ref.\cite{RR}.

We define the "small" dimensionless quantities $\alpha =\mu/m$, $ \kappa = |\mbox{\boldmath $\q$}|/m$, and use the standard techniques for loop 
integration in order to write
\begin{eqnarray}
\hat{J}_{c,c}(m; \mbox{\boldmath $\q$}) &=& - \int_0^1 da \; \ln\left[ 1+ a(1\!-\!a)\; \kappa^2/ \alpha^2\right] 
\nonumber\\[2mm]
&=&  2- \sqrt{1+ 4\; \alpha^2/\kappa^2}\; \ln\left[ \frac{ \sqrt{1+ 4 \;\alpha^2/\kappa^2}+1}{ \sqrt{1+ 4 \;\alpha^2/\kappa^2}-1}\right] \;,
\label{9}\\[2mm]
\hat{J}_{c,sN}^{(1)}(m; \mbox{\boldmath $\q$}) &=& 2 \int_0^1 da \;(1\!-\!a)^2 \int_0^1 db\; (1\!-\!b)
\nonumber\\[2mm]
&\times& \left\{ \left[ a(1\!-\!a)b\; \kappa^2 + (a+b-ab)\; \alpha^2 + (1\!-\!a)^2 (1\!-\!b)^2 \right]^{-1} \right.
\nonumber\\[2mm]
&-& \left.\left[ (a+b-ab)\; \alpha^2 + (1\!-\!a)^2 (1\!-\!b)^2\right]^{-1} \right\} 
\nonumber\\[2mm]
&=& \hat{J}_{c,c}(m; \mbox{\boldmath $\q$}) - \left(\frac{\pi\mu}{m}\right)\; \hat{I}_N(m; \mbox{\boldmath $\q$}) \;,
\label{10}
\end{eqnarray}
where the integral $\hat{I}_N(m; \mbox{\boldmath $\q$})$ is given by
\begin{eqnarray}
&&\hat{I}_N(m; \mbox{\boldmath $\q$}) = -\left(\frac{m}{\pi\mu}\right)
\int_0^1 da \left\{ \frac{ (\alpha^2+a\; \kappa^2)}{\sqrt{\alpha^2+a(1\!-\!a)\;\kappa^2 - (\alpha^2+a \;\kappa^2)^2 /4}}\right.
\nonumber\\[2mm]
&&\times \left.\left[ \tan^{-1}\left( \frac{(1\!-\!a) - (\alpha^2+a\;\kappa^2)/2}{\sqrt{\alpha^2+a(1\!-\!a)\;\kappa^2 -(\alpha^2+a\;\kappa^2)^2 /4}} \right) \right.\right.
\nonumber\\[2mm]
&&+ \left.\left. \tan^{-1}\left( \frac{ (\alpha^2+a\;\kappa^2)/2}{\sqrt{\alpha^2+a(1\!-\!a)\;\kappa^2 -(\alpha^2+a\;\kappa^2)^2/4}}\right) \right] \right.
\nonumber\\[2mm]
&&- \left.\frac{\alpha^2}{\sqrt{\alpha^2 - \alpha^4 /4}}
\left[ \tan^{-1}\left( \frac{(1\!-\!a) - \alpha^2/2}{\sqrt{\alpha^2 - \alpha^4/4}} \right) 
+  \tan^{-1}\left( \frac{ \alpha^2/2}{\sqrt{\alpha^2 - \alpha^2/4}}\right) \right] \right\} \;.
\label{11}
\end{eqnarray}

The function $\sigma_N(t)$ and the asymptotic central potential can then be written simply as 
\begin{eqnarray}
\hat{V}_a^S(m; \mbox{\boldmath $\q$}) 
&=& -\;\frac{2}{f_\pi^2\mu^2} 
\;\left[ f_\pi^2 \left( a_{00}^+ + t \;a_{01}^+ \right) \right] \; \sigma_N(t)
\nonumber\\[2mm]
&=& - \;\frac{2}{f_\pi^2\mu^2}\;
\left[ f_\pi^2 \left( a_{00}^+ -\mbox{\boldmath $\q$}^2 \;a_{01}^+ \right) \right]
\left[ \frac{3 \mu^3}{32\pi}\left( \frac{g}{m}\right)^2 \hat{I}_N(m; \mbox{\boldmath $\q$})\right] \;.
\label{12}
\end{eqnarray}

The first result is very general and independent of specific models or approximation schemes used to calculate the 
scalar form factor.
It has a rather transparent physical interpretation: 
the factor proportional to $\sigma_N(t)$ represents, as one will show in the sequence, 
the pion cloud of one of the nucleons, which is extended in space;
the term involving the subthreshold coefficients in the square bracket represents the other nucleon, 
acting as a scalar source expanded in multipoles; 
the factor 2 arises because the potential is due to the crosed term in $(N+R)^2$
and the scale $(-f_\pi^2\mu^2)$ corresponds to the energy density of the quark condensate.

\begin{table}[h]
\begin{center}

\caption{Values for the subthreshold coefficients $a_{00}^+$, $a_{01}^+$, and the scalar form factor at the Cheng-Dashen point, $\sigma(t=2\mu^2)$. Labels (emp), $(s)$, and $(\Delta_\ell)$ denote respectively empirical results, 
theoretical scalar form factor and leading delta contributions.}

\begin{tabular} {|c||c|c|c||c|c|c|}
\hline
& emp \cite{H83}& emp \cite{KH}& emp \cite{PA}& $\;\;\;\;\;s\;\;\;\;\;$ & $\;\;\;\;\;\Delta_\ell\;\;\;\;\;$ & $\;s+\Delta_\ell\;$ \\ \hline\hline
$ a_{00}^+\;\;( \mu^{-1}) $ & $-1.46\pm0.10 $ &$-1.30\pm0.02$  & $-1.27\pm0.03$  & 0.72 & -1.89 & -1.20  \\ \hline
$ a_{01}^+\;\;( \mu^{-3}) $ & $1.14\pm0.02$   &$1.35\pm0.14 $  & $1.27\pm0.03 $  & 0.12 & 0.94  & 1.12  \\ \hline
$ \sigma(2\mu^2) $ (MeV)\;\; & $ 64\pm8 $&$88\pm15$ & $90\pm8 $ & & & \\ \hline
\end{tabular}

\label{Tab.1}

\end{center}
\end{table}

Empirical  values of the subthreshold coefficients can be obtained by means of dispersion relations and some results
available in the litterature are given in table \ref{Tab.1}.
As pointed out long ago\cite{OO}, they may be well represented by a model including the delta and supplemented by 
information about the scalar form factor.
In order to use this information, one writes $\sigma(t) \cong \sigma+t\;\sigma'$, where $\sigma$ is the 
$\pi N$ $\sigma$-term.
The parameter $\sigma'$, the slope at the origin, may also be obtained by means of dispersion relations.
For instance, a recent analysis \cite{GLS}, based on $\sigma(2\mu^2)=60$ MeV, has produced $\sigma=45$ MeV, 
which corresponds to $\sigma'=0.054 \mu^{-1}$.
The scalar form factor is then related to the subthreshold coefficients by $a_{00}^+)_s=\sigma/f_\pi^2$ and
$a_{01}^+)_s=\sigma'/f_\pi^2$, where $f_\pi$ is the pion decay constant.

As far as the delta is concerned, one uses the results of ref.\cite{CDR} and obtains 
\begin{eqnarray}
a_{00}^+)_\Delta &=& - \;g_\Delta^2 \left(\frac{8}{9}\right)\; \frac{\mu^2 (m+M_\Delta)}{M_\Delta^2-m^2-\mu^2}\;
\left( 1-\;\frac{\mu^2}{M_\Delta^2}\right) \;,
\label{13}\\[2mm]
a_{01}^+)_\Delta &=&  \;g_\Delta^2 \left(\frac{4}{9}\right)\; \frac{(m+M_\Delta)}{M_\Delta^2-m^2-\mu^2}\;
\left\{ \left( 1-\;\frac{\mu^2}{M_\Delta^2}\right) \right.
\nonumber\\[2mm]
&+& \left. \frac{\mu^2}{2 M_\Delta^2 (M_\Delta^2-m^2-\mu^2)}
\left[ M_\Delta(3 M_\Delta-m) - \mu^2 \frac{(3 M_\Delta+2m)}{(M_\Delta+m)}\right] \right\}\;,
\label{14}
\end{eqnarray}
where $M_\Delta$ is the delta mass, $g_\Delta$ is the pion-nucleon-delta coupling constant and we have used $Z=-1/2$.
The constant $g_\Delta$ can be related to $g$ by means of the large $N_c$ result $2m g_\Delta = 3 g /\sqrt{2}$ \cite{BKM}
and the leading delta contributions may be written as
\begin{eqnarray}
a_{00}^+)_\Delta &=& - \left(\frac{g}{m}\right)^2 \; \frac{\mu^2}{M_\Delta-m} \;,
\label{15}\\[2mm]
a_{01}^+)_\Delta &=&  \frac{1}{2}\; \left(\frac{g}{m}\right)^2 \; \frac{1}{M_\Delta-m}\;.
\label{16}
\end{eqnarray}
Adopting $g=13.40$ and $f_\pi=93$ MeV, one has the values displayed in table 1.

Most of the dependence of the potential on the momentum transferred is embodied on the dimensionless function 
$\hat{I}_N(m; \mbox{\boldmath $\q$})$, which is also influenced by the value of the nucleon mass.
In the heavy baryon (HB) limit, defined as $m\rightarrow\infty$, this function can be evaluated analytically and one has 
\cite{KBW}
\begin{eqnarray}
\hat{I}_N(m\rightarrow\infty; \mbox{\boldmath $\q$}) \equiv \hat{I}_{HB}(\mbox{\boldmath $\q$}) &=& -\left(\frac{m}{\pi\mu}\right) \frac{\pi}{2}
\int_0^1 da \left[ \frac{u^2+a q^2}{\sqrt{u^2+a(1\!-\!a)\;q^2}}-u \right]
\nonumber\\[2mm]
&=& - \left[ \frac{\mu^2+\mbox{\boldmath $\q$}^2/2}{\mu\; |\mbox{\boldmath $\q$}|}\;\tan^{-1}\frac{|\mbox{\boldmath $\q$}|}{2 \mu} -\frac{1}{2} \right] \;.
\label{17}
\end{eqnarray}

The expression for the heavy baryon potential in momentum space is then
\begin{equation}
\hat{V}_{HB}(\mbox{\boldmath $\q$}) = \frac{3\mu}{16\pi}\left( \frac{g_A}{f_\pi}\right)^2
\left( a_{00}^+ - \mbox{\boldmath $\q$}^2 a_{01}^+ \right)
\left[ \left( \frac{\mu^2+\mbox{\boldmath $\q$}^2/2}{\mu\; |\mbox{\boldmath $\q$}|}\right)\;\tan^{-1}
\left( \frac{|\mbox{\boldmath $\q$}|}{2\mu}\right)  - \;\frac{1}{2}\right]\;,
\label{18}
\end{equation}
where we have used the Goldberger-Treiman relation $g/m=g_A/f_\pi$.
This result reproduces, up to irrelevant polynomial terms, the potential derived in ref.\cite{KBW} using HBChPT, 
when one identifies the subthreshold coefficients and the low energy constants $c_i$ as follows: 
$a_{00}^+ = -\mu^2(4c_1-2c_3)/f_\pi^2$ and $a_{01}^+ = -c_3/f_\pi^2$.
It is worth pointing out that the numerical values recommended in that work yield $a_{00}^+ = -2.65 \;\mu^{-1}$ and
$a_{01}^+ =1.66\; \mu^{-3}$, which deviate from those quoted in table 1.
This happens because the constants $c_i$ include both leading terms and first order corrections\cite{BKM}.
The $c_i$ then take the latter into the potential and give rise to fourth order corrections to otherwise third order expressions. 
When these corrections are eliminated for the sake of a consistent power counting, one recovers values compatible with table 1.

Heavy baryon chiral perturbation theory was also used in other calculations \cite{ORK}, based on time ordered 
perturbation theory. 
In order to allow comparison with our results, we perform Cauchy integrations in eqs.(\ref{6}) 
and (\ref{7}) and obtain
\begin{eqnarray}
&&\hat{I}_N(m; \mbox{\boldmath $\q$}) = \frac{m}{\pi \mu} \left( \hat{J}_{c,c}-\hat{J}_{c,sN}^{(1)} \right)
= \frac{4\pi m}{\mu}\int \frac{d^3\mbox{\boldmath $Q$}}{(2\pi)^3}
\left\{ \frac{1}{\omega_+\omega_-(\omega_+ \!+\!\omega_-)}\right.
\nonumber\\[2mm]
&& \left. -\; \frac{2m}{V^2}\left[ \frac{(mV^0+E_Q)V^0+\mbox{\boldmath $V$}\!\cdot\!\mbox{\boldmath $Q$}}
{E_Q [(mV^0+E_Q)^2\!-\!\omega_+^2] [(mV^0+E_Q)^2\!-\!\omega_-^2]}
\right.\right.
\nonumber\\[2mm]
&& \left.\left.+ \frac{1}{(\omega_+^2-\omega_-^2)}
\left( \frac{\omega_+V^0+\mbox{\boldmath $V$}\!\cdot\!\mbox{\boldmath $Q$}}{\omega_+\;[(mV^0-\omega_+)^2-E_Q^2]}
- \frac{\omega_-V^0+\mbox{\boldmath $V$}\!\cdot\!\mbox{\boldmath $Q$}}{\omega_-\;[(mV^0-\omega_-)^2-E_Q^2]}\right) \right]\right.
\nonumber\\[2mm]
&&\left. -\left(\omega_+\rightarrow\sqrt{\mu^2+\mbox{\boldmath $Q$}^2},\;\; \omega_- \rightarrow\sqrt{\mu^2+\mbox{\boldmath $Q$}^2}\right)  \right\} \;,
\label{19}
\end{eqnarray}
where $\omega_{\pm} = \sqrt{\mu^2+(\mbox{\boldmath $Q$}^2\pm \mbox{\boldmath $\q$}^2/4)}$ and
$E_Q =\sqrt{m^2+(\mbox{\boldmath $Q$}+m\mbox{\boldmath $V$})^2}$.

Going to the limit $m\rightarrow\infty$, one has
\begin{equation}
\hat{I}_{HB}(\mbox{\boldmath $\q$}) = \frac{4\pi}{\mu}\int \frac{d^3\mbox{\boldmath $Q$}}{(2\pi)^3}
\left[ \frac{\mbox{\boldmath $Q$}^2-\mbox{\boldmath $\q$}^2/4}{\omega_+^2 \; \omega_-^2}
-\;\frac{\mbox{\boldmath $Q$}^2}{(\mu^2+\mbox{\boldmath $Q$}^2)^2}\right] 
\label{20}
\end{equation}
and a potential given by 
\begin{equation}
\hat{V}_{HB}(\mbox{\boldmath $\q$}) = -\;\frac{3}{4}\left( \frac{g}{m}\right)^2
\left(a_{00}^+ -\mbox{\boldmath $\q$}^2 a_{01}^+\right)
\int \frac{d^3\mbox{\boldmath $Q$}}{(2\pi)^3} \left[ \frac{\mbox{\boldmath $Q$}^2-\mbox{\boldmath $\q$}^2/4}{\omega_+^2 \; \omega_-^2}
-\;\frac{\mbox{\boldmath $Q$}^2}{(\mu^2+\mbox{\boldmath $Q$}^2)^2}\right] \;.
\label{21}
\end{equation}
\begin{figure}[h]
\begin{center}
\includegraphics[width=15cm]{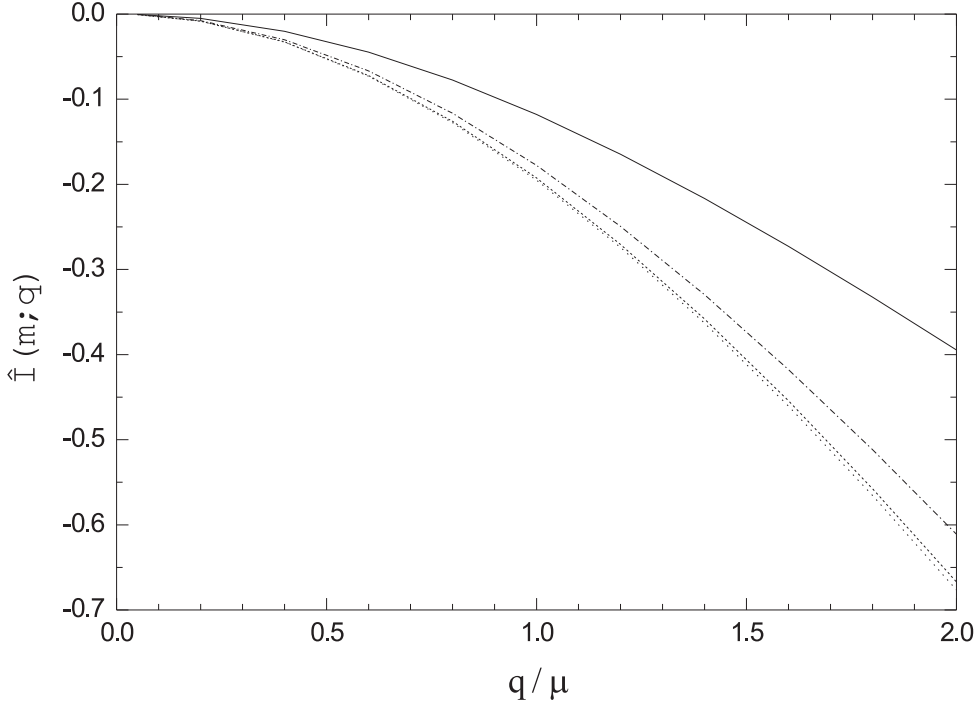}
\caption{Dependence of the function $\hat{I}_N(m;\mbox{\boldmath $\q$})$, eq.(\ref{11}), on the nucleon mass, for 
$m=m_N$ (continuous line),  $m=10\;m_N$ (dot-dashed line), $m=100\;m_N$ (dashed line), $m=1000\;m_N$ (dotted line),
where $m_N$ is the empirical nucleon mass.}
\label{fig.2}
\end{center}
\end{figure}
\begin{figure}[h]
\begin{center}
\includegraphics[width=15cm]{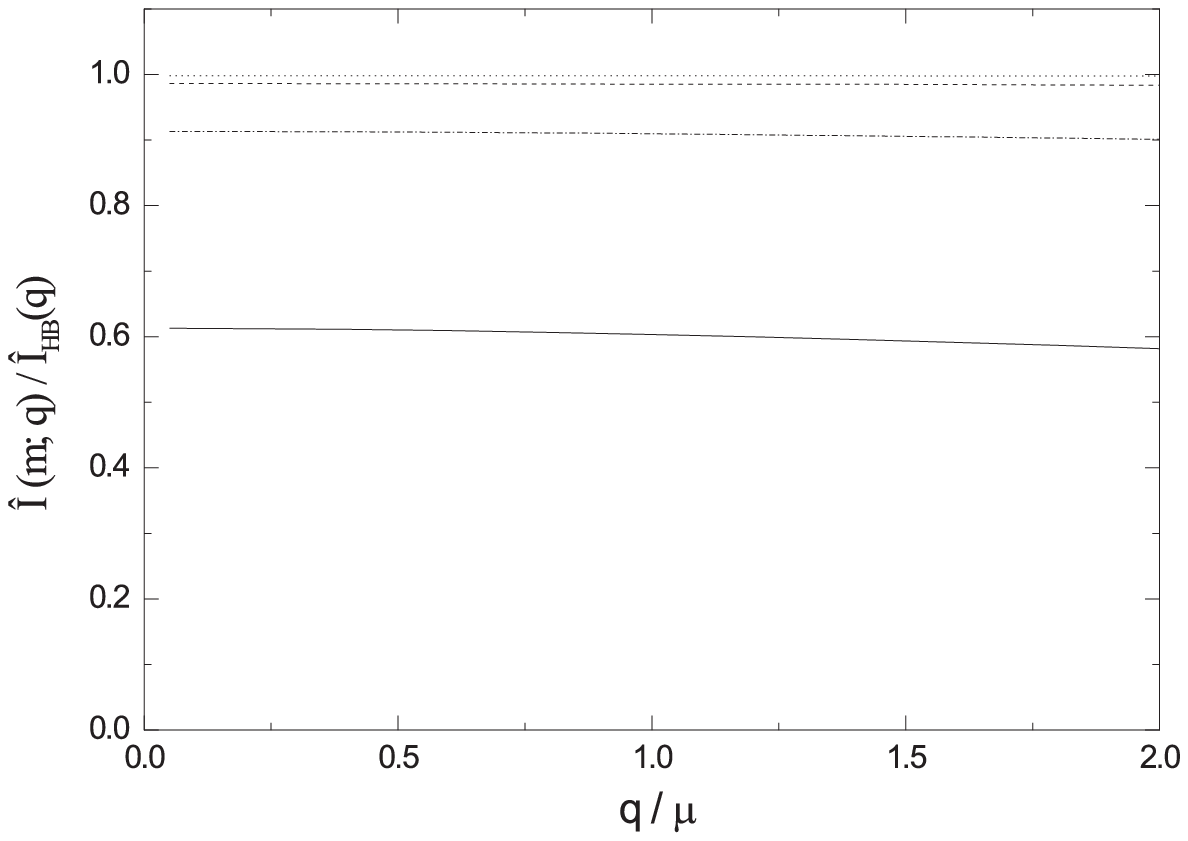}
\caption{Dependence of the ratios of the functions $\hat{I}_N(m;\mbox{\boldmath $\q$})$, eq.(\ref{11}) 
and $\hat{I}_{HB}(\mbox{\boldmath $\q$})$, eq.(\ref{17}), on the nucleon mass; conventions as in fig.\ref{fig.2}.}
\label{fig.3}
\end{center}
\end{figure}

In order to check the consistency of this result, we have integrated analytically eq.(\ref{20}) and recovered 
eq.(\ref{17}) exactly.

In fig.\ref{fig.2} we display the behaviour of $\hat{I}_N(m; \mbox{\boldmath $\q$})$ for $m = m_N, 10\; m_N, 100 \;m_N$ 
and $1000\; m_N$, where $m_N=938.28$ MeV is the empirical nucleon mass.
The ratios $\hat{I}_N(m; \mbox{\boldmath $\q$}) / \hat{I}_{HB}(\mbox{\boldmath $\q$})$ for the same values of $m$ are shown in fig.\ref{fig.3}, where it is possible to see that typical deviations from the heavy baryon limit are respectively
about 30\%, 10\%, 2\% and less than $1\%$.
Thus, one learns that the heavy baryon limit is about 30\% far from the real world.
As we discuss in the next section, this is related to the fact that the heavy baryon expansions of loop integrals sometimes fail to 
converge \cite{BL}.

The potential in coordinate space is given by 
\begin{equation}
\hat{V}_a^S(m; \mbox{\boldmath $r$}) = 
- \;\frac{2}{f_\pi^2\mu^2}\;\left[ f_\pi^2 \left( a_{00}^+ +\mbox{\boldmath $\nabla$}^2 \;a_{01}^+ \right) \right]\;
\sigma_N(m; \mbox{\boldmath $r$})  \;,
\label{22}
\end{equation}
where $\sigma_N(m; \mbox{\boldmath $r$})$ is the form factor in configuration space.
It is convenient to write it in terms of a dimensionless function $\hat{F}_N(m; \mbox{\boldmath $r$}) $ as 
\begin{equation}
\sigma_N(m; \mbox{\boldmath $r$}) =
 \frac{3 \mu^6}{128\pi^2}\left( \frac{g}{m}\right)^2 \hat{F}_N(m; \mbox{\boldmath $r$}) \;,
\label{23}
\end{equation}
with  
\begin{equation}
\hat{F}_N(m; \mbox{\boldmath $r$}) = \frac{4\pi}{\mu^3}\;\int \frac{d^3\mbox{\boldmath $\q$}}{(2\pi)^3}\; e^{-i\mbox{\boldmath $\q$}\cdot \mbox{\boldmath $r$}}\;\hat{I}_N(m; \mbox{\boldmath $\q$}) \;.
\label{24}
\end{equation}

In the heavy baryon limit this expression can be evaluated explicitly \cite{KBW} and, using $x=\mu r$, one writes
\begin{eqnarray}
\hat{F}_N(m\rightarrow\infty; \mbox{\boldmath $r$}) \equiv  \hat{F}_{HB}(\mbox{\boldmath $r$}) &=&
\left( 1+\frac{2}{x}+\frac{1}{x^2}\right)\; \frac{e^{-2x}}{x^2}
= \left[ \left( 1+\frac{1}{x}\right) \frac{e^{-x}}{x} \right]^2 \;,
\label{25}\\[2mm]
\mbox{\boldmath $\nabla$}_{\!x}^2 \hat{F}_{HB}(\mbox{\boldmath $r$})&=& 
4 \left( 1+\frac{3}{x}+\frac{11}{2\;x^2}+\frac{6}{x^3}+\frac{3}{x^4}\right) \frac{e^{-2x}}{x^2} \;.
\label{26}
\end{eqnarray}

For generic values of $m$, it is more convenient to perform the Fourier transform using directly 
eqs.(\ref{9}) and (\ref{10}) and one has \cite{MPR}
\begin{equation}
\left( \frac{t}{\mu^2}\right)^p\hat{F}_N(m; \mbox{\boldmath $r$}) = \frac{m}{\pi \mu}\; \int_0^1 da \int_0^1 db 
\left[ \frac{\lambda^{2p+2}}{b}\; \frac{e^{-\lambda x}}{x} 
-\; \frac{2m^2}{\mu^2}\;\eta^{2p}\; \frac{(1\!-\!a)(1\!-\!b)}{a\;b}\; \frac{e^{-\eta x}}{x}\right] \;,
\label{27}
\end{equation}
where 
\begin{eqnarray}
\lambda^2 &=& 1 / \left[ a(1-a) b\right] \;,
\label{28}\\[2mm]
\eta^2 &=& \left[ a+b-ab + (1\!-\!a)^2 (1\!-\!b)^2 \;m^2/ \mu^2 \right] \;\lambda^2 \,.
\label{29}
\end{eqnarray}

\begin{figure}[h]
\begin{center}
\includegraphics[width=15cm]{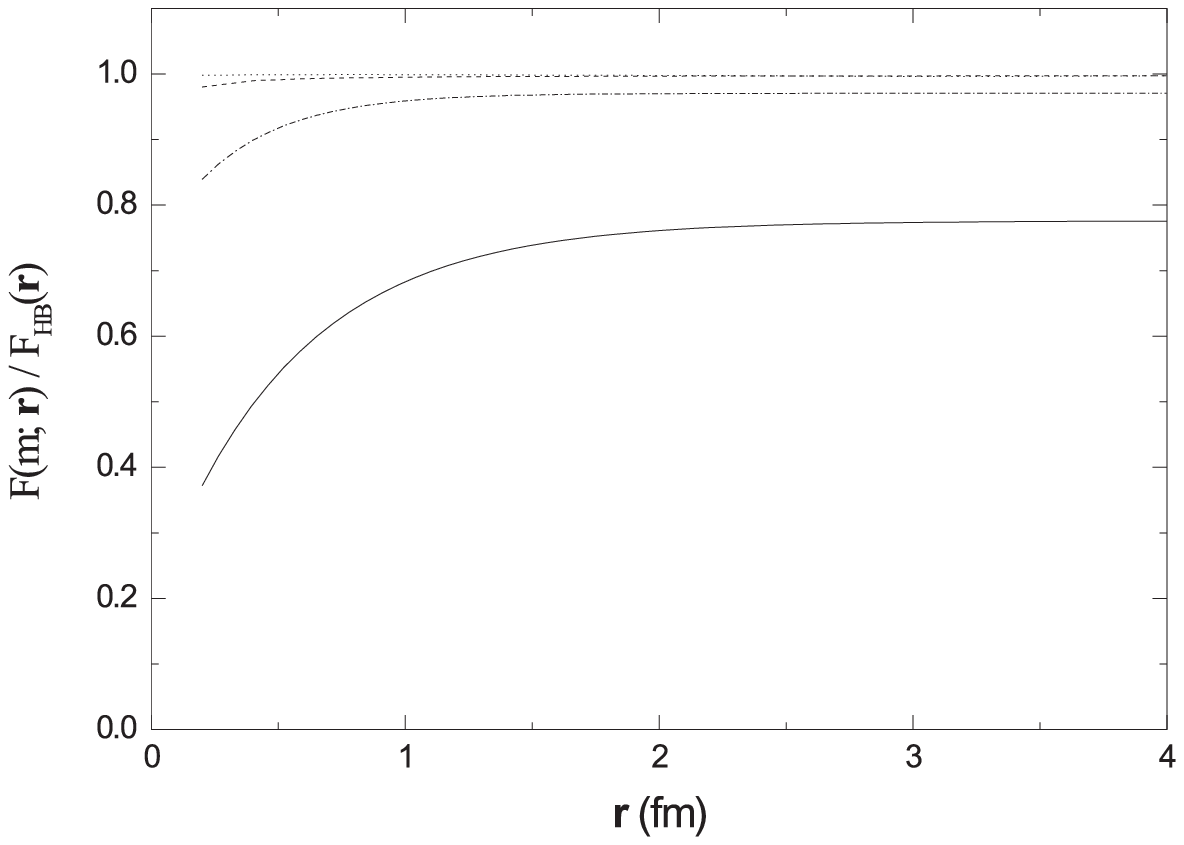}
\caption{Dependence of the ratios of the functions $\hat{F}_N(m;\mbox{\boldmath $r$})$, eq.(\ref{27}) and $\hat{F}_{HB}(\mbox{\boldmath $r$})$, eq.(\ref{25}), on the nucleon mass; conventions as in fig.\ref{fig.2}.}
\label{fig.4}
\end{center}
\end{figure}

The behaviour of the ratio $\hat{F}_N(m; \mbox{\boldmath $r$})/\hat{F}_{HB}(\mbox{\boldmath $r$})$ as function of $m$ is displayed in fig.\ref{fig.4}, where one finds the same qualitative pattern as in figs.\ref{fig.2} and \ref{fig.3}.
In configuration space, however, the gap between the heavy baryon limit and results with the empirical value of the nucleon mass
varies with distance and is about 25\% asymptotically.
The difference between the sizes of the gaps in momentum and configuration spaces can be understood by noting that 
the former may contain pure polynomials, which correspond to zero-range functions and hence are not present in the latter. 

\section{Chiral Symmetry}

The behaviour of the triangle integral that determines the scalar form factor has been extensively discussed in a recent work 
by Becher and Leutwyler \cite{BL}.
The leading term in their chiral expansion of $\sigma(t)$ can be obtained by neglecting terms proportional to $t/4m^2$ in
eq.(\ref{27}) and one has 
\begin{equation}
\hat{F}_\ell(m; \mbox{\boldmath $r$}) = -\; \frac{m}{\pi\mu}\; \int_0^1 da \int_0^1 db \;
\frac{(1- \eta^2/2 )}{a\;b}\; \frac{e^{-\eta x}}{x}\;.
\label{30}
\end{equation}

This result is equivalent to the dispersive representation of ref.\cite{BL}, given by 
\begin{equation}
\hat{F}_\ell (m; \mbox{\boldmath $r$}) = -\;\frac{1}{\pi\mu^3}
\int_{4\mu^2}^\infty dt\;\frac{(\mu^2-t/2)}{\sqrt{t}}\;G_\ell(m; t, \mu) \;\frac{e^{-\sqrt{t}x/\mu}}{x} \;.
\label{31}
\end{equation}
with
\begin{equation}
G_\ell (m; t ,\mu ) = \tan^{-1}\left[ \frac{2m \sqrt{t-4\mu^2}}{t-2\mu^2}\right] \;.
\label{32}
\end{equation}

In order to study the structure of this integral, we rewrite it as 
\begin{equation}
\hat{F}_\ell (m; \mbox{\boldmath $r$}) = -\;\frac{1}{\pi\mu^3}\;\frac{e^{-2x}}{x}\;
\int_{4\mu^2}^\infty dt\;\frac{(\mu^2-t/2)}{\sqrt{t}}\;G_\ell (m; t ,\mu ) \;e^{-(\sqrt{t}-2\mu) x/\mu} \;.
\label{33}
\end{equation}

For large values of $x$, the exponential in the integrand falls off very rapidly as a function of $\sqrt{t}$ and the 
integral tends to be dominated by a small region around $t=4\mu^2$.
On the other hand, the function $G_\ell$ vanishes at this point and hence the net result is due to a compromise 
between these two effects.
In the work of Becher and Leutwyler one learns that it is possible to write accurately
\begin{equation}
G_\ell(m; t,\mu) \cong \left[\frac{\pi}{2}\right]_{\!HB}
-\left[ \frac{\alpha(t-2\mu^2)}{2\sqrt{t\!-\!4\mu^2}}\right]_{\!A}
+ \left[ \frac{\alpha\sqrt{t}}{2\sqrt{t\!-\!4\mu^2}}
- \frac{\sqrt{t}}{2\mu} \tan^{-1}\frac{\alpha\mu}{\sqrt{t\!-\!4\mu^2}}\right]_{\!B}\;,
\label{34}
\end{equation}  
where $\alpha = \mu/m$.
The first two terms in this expression correspond to the standard heavy baryon series whereas the last one implements
the correct analytic behaviour around the point $t=4\mu^2$.
In particular, the $HB$ term gives rise to eq.(\ref{25}) and has been used in HBChPT calculations of the $NN$ potential.
\begin{figure}[h]
\begin{center}
\includegraphics[width=15cm]{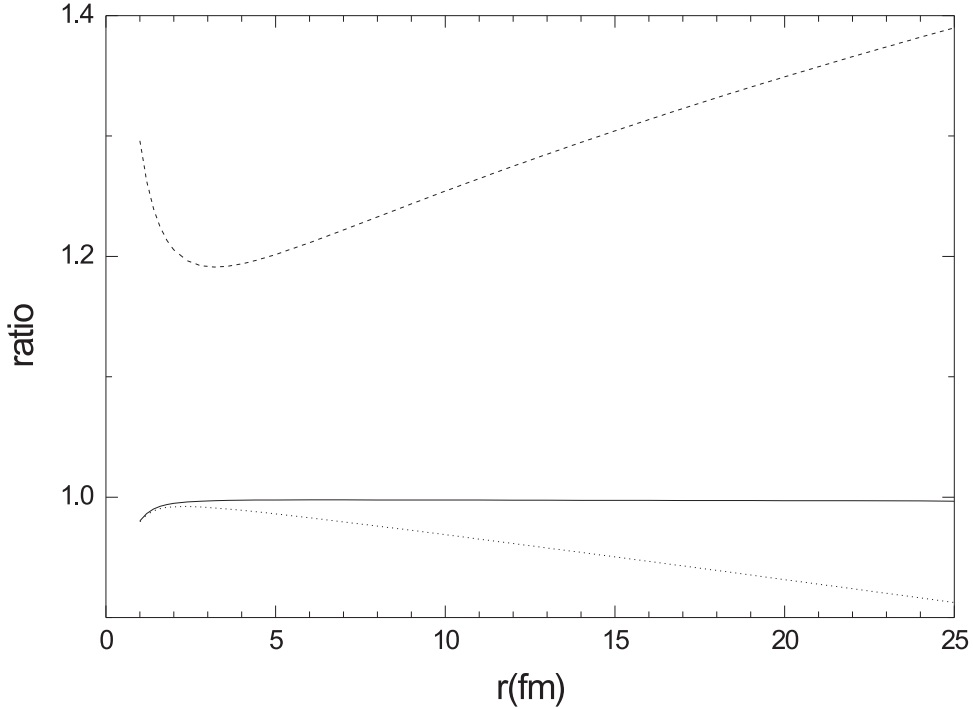}
\caption{Ratios of the functions $\hat{F}_{HB}$ (dashed line), $\hat{F}_{HB}+\hat{F}_A$ (dotted line) 
and $\hat{F}_{HB}+\hat{F}_A +\hat{F}_B$ (continuous line) , obtained by using eq.(\ref{34}) into eq.(\ref{31}), 
by $\hat{F}_\ell$, eqs.(\ref{30}-\ref{32}).}
\label{fig.5}
\end{center}
\end{figure}

In fig.\ref{fig.5} we display the ratios of the functions $\hat{F}_{HB}$, $\hat{F}_{HB}+\hat{F}_A$ and
$\hat{F}_{HB}+\hat{F}_A +\hat{F}_B$, obtained by
using eq.(\ref{34}) into eq.(\ref{31}), by $\hat{F}_\ell$.
The first thing to be noted is that there is a gap between $\hat{F}_{HB}$ and $\hat{F}_\ell$, which is 
about $20\%$ in the region of physical interest and can be as large as $40\%$ asymptotically.
The inclusion of $\hat{F}_A$ brings the approximate result much closer to the exact one, indicating that the 
last term in eq.(\ref{34}) is important mainly at rather large distances.
Thus, in principle, HBChPT could allow a precise numerical description of the $NN$ potential,
but this would require the extension of existing calculations to higher orders. 
This illustates the advantage of the relativistic framework proposed recently by Becher and Leutwyler, 
namely that it gives rise automatically to leading contributions which are already rather accurate.
 
The leading chiral contribution to the central potential is obtained by using eqs.(\ref{31}) and (\ref{32}) into 
eqs.(\ref{23}) and (\ref{22}).
In order to assess the phenomenological implications of  this result, one could evaluate phase shifts and compare them 
with experiment. 
However, one should bear in mind that, in contributing to phase shifts, the central TPEP is superimposed to both isospin
dependent components and a large OPEP background.
Its effects are completely blurred in the $^1S_0$ wave, but may be identified in the waves $^1D_2$ and $^1G_4$.
In this situation, it becomes possible for different chiral models to accomodate experimental data\cite{ORK,BRR,KBW,KGW}.
In fact, in the case of the TPEP, phase shifts do not represent the most stringent test possible. 
\begin{figure}[h]
\begin{center}
\includegraphics[width=15cm]{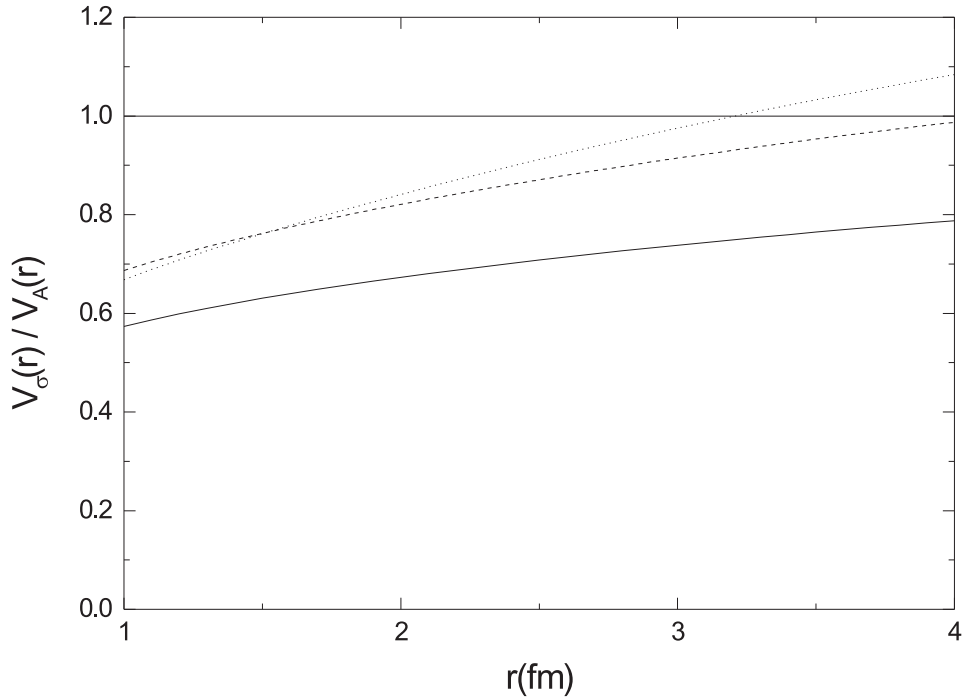}
\caption{Ratio of the leading asymptotic relativistic chiral prediction 
$\hat{V}_\sigma (\mbox{\boldmath $r$})$, eqs.(\protect\ref{22}, \protect\ref{23}) amd (\protect\ref{30}-\protect\ref{32}), 
by the central medium-range Argonne potential \protect\cite{Arg},
for the values of the subthreshold coefficients of refs. 
\protect\cite{H83} (continuous line), \protect\cite{KH} (dashed line) and \protect\cite{PA} (dotted line).}
\label{fig.6}
\end{center}
\end{figure}

Therefore, in this work we compare the predictions of the asymptotic $\hat{V}_a^S(m; \mbox{\boldmath $r$}) $ with the
Argonne phenomenological potential\cite{Arg}, which is rather accurate and constrained by a large data base, that also
includes information about the $^1S_0$ channel.
In fig.\ref{fig.6} we plot the ratio of the leading chiral potential by the corresponding Argonne component, 
in the range 1fm$\leq r \leq$4fm, for the sets of subthreshold coefficients considered in table 1.
Typical discrepancies are of the order of 20\%, indicating that the very simple mechanism described in the previous section is indeed responsible for the leading contribution to the central potential.
The existing gap has the order of magnitude expected from chiral corrections.
It is worth pointing that, in the case of $NN$ interactions, these corrections would affect not only the scalar form factor, 
but also require the evaluation of other amplitudes.
In particular, in the case of the delta contributions, one has also to include box and crossed box diagrams.
A comprehensive discussion of this problem will therefore be postponed.

\section{Other Results}

We begin by considering first order corrections to the leading result for $\sigma(t)$, which come from two different sources.
One of them corresponds to the terms neglected when passing from eq.(\ref{27}) to eq.(\ref{30}), whereas the other 
one is implemented by subthreshold coefficients, representing the remainder $R$ discussed in the previous section.
Thus, the scalar form factor, including both leading and first order corrections, is given by 
\begin{equation}
\sigma(m; \mbox{\boldmath $r$}) =\sigma_N(m; \mbox{\boldmath $r$}) +\sigma_R(m; \mbox{\boldmath $r$})\;,
\label{35}
\end{equation}
where $\sigma_N$ is the result of ref.\cite{GSS}, given by eq.(\ref{29}) and $\sigma_r$ is written in 
terms of the subthreshold coefficients $d_{ij}^+$ as\cite{BL}
\begin{equation}
\sigma_r(m; \mbox{\boldmath $r$}) = \frac{3\mu^4}{128\pi^2}
\left\{ \frac{\mu^2}{m}
\left[ \left( d_{00}^+ +t\; d_{01}^+ \right) +\frac{1}{3}\left(\mu^2 - t/4\right) d_{10}^+ \right] \right\}\;
\hat{F}_{cc} (\mbox{\boldmath $r$}) \;,
\label{36}
\end{equation}
with
\begin{equation}
\hat{F}_{cc}(\mbox{\boldmath $r$}) = \frac{4m}{\mu^4}\;\int \frac{d^3\mbox{\boldmath $\q$}}{(2\pi)^3}\; e^{-i\mbox{\boldmath $\q$}\cdot \mbox{\boldmath $r$}}\;\hat{J}_{cc}(\mbox{\boldmath $\q$}) \;.
\label{37}
\end{equation}

The main contributions to the coefficients entering this result come from the $\sigma$-term and the $\Delta$ 
and one may also write
\begin{equation}
\sigma_r(m; \mbox{\boldmath $r$}) \approx
\sigma_{\sigma+\Delta}(m; \mbox{\boldmath $r$})  =
\frac{3\mu^4}{128\pi^2}\;\frac{\mu^2\sigma}{m f_\pi^2}  \;\hat{F}_{cc}(\mbox{\boldmath $r$}) +
\sigma_\Delta(m; \mbox{\boldmath $r$}) \;.
\label{38}
\end{equation}
The term $\sigma_\Delta$ is given by the diagram of fig.\ref{fig.1}b.
Using the results of ref.\cite{CDR} one finds, after neglecting terms ${\cal O}(t/m^2)$
\begin{equation}
\sigma_\Delta(m; \mbox{\boldmath $r$}) =
 \frac{3 \mu^6}{128\pi^2}\left( \frac{g}{m}\right)^2 
\frac{(m+M_\Delta)}{m}\; \hat{F}_\Delta (m; \mbox{\boldmath $r$})  \;,
\label{39}
\end{equation}
where $\hat{F}_\Delta$ is obtained from the $\hat{F}_N$ of eq.(\ref{30}) by replacing $\eta^2$ with
\begin{equation}
\theta^2=\eta^2 + [(1\!-\!a) (1\!-\!b)\; (M_\Delta^2-m^2)/\mu^2]\; \lambda^2 \;.
\label{40}
\end{equation}
\begin{figure}[h]
\begin{center}
\includegraphics[width=15cm]{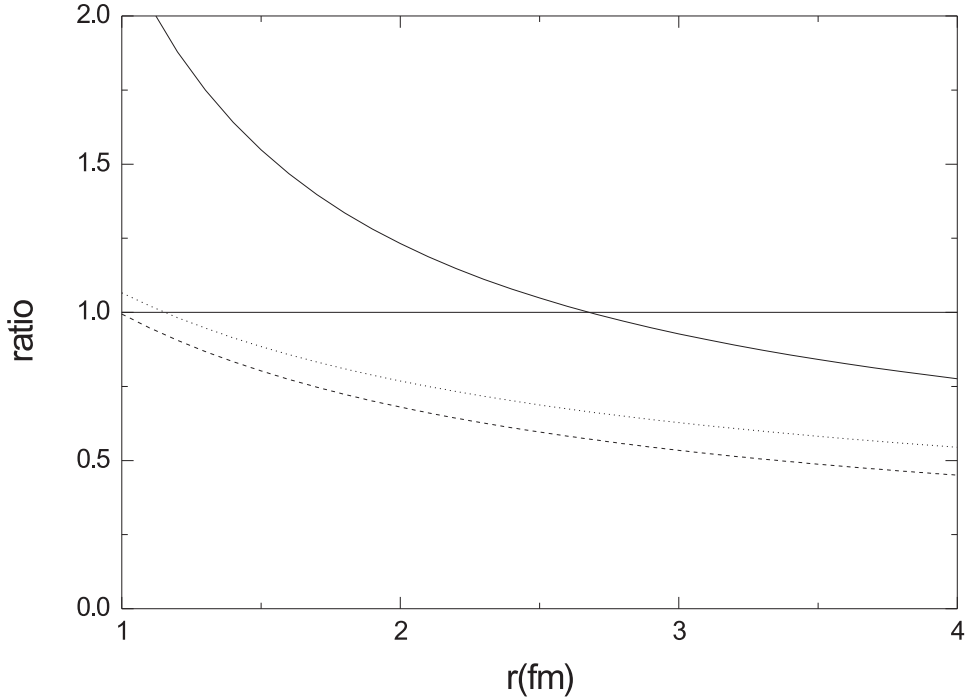}
\caption{Ratios of $\sigma_R$ eq.(\ref{36} - continuous line), $\sigma_\Delta$ eq.(\ref{39} - dashed line)  and $\sigma_{s+\Delta}$ eq.(\ref{38} - dotted line) by the leading chiral contribution 
$\sigma_\ell$ eqs.(\ref{30}-\ref{32}) and (\ref{23}).}
\label{fig.7}
\end{center}
\end{figure}

As the function $\theta$ determines the effective mass of the two-pion system exchanged between the nucleon and the  scalar source, this result indicates that $\sigma_\Delta$ has shorter range than $\sigma_N$.
In fig.\ref{fig.7}, we display the ratios of $\sigma_r$, $\sigma_\Delta$ and $\sigma_{s+\Delta}$ by the leading chiral contribution
$\sigma_\ell$ and one notes that these corrections remain quite visible at relatively large distances.

\begin{figure}[h]
\begin{center}
\includegraphics[width=15cm]{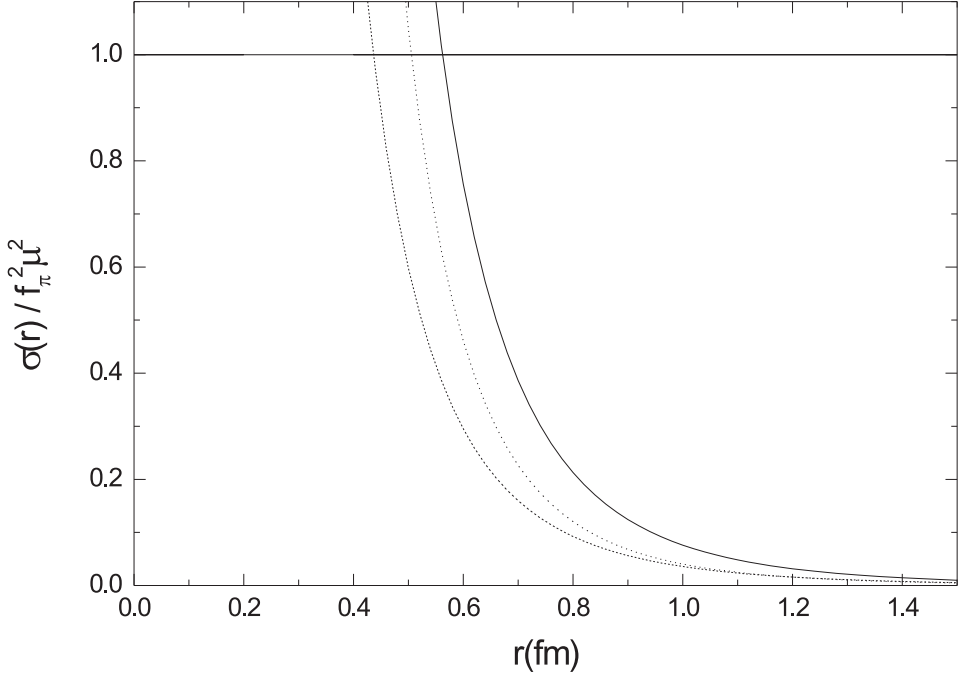}
\caption{Spatial dependence of the scalar form factor, normalized by the quark condensate density:
$\sigma_N$ eq.(\ref{23} - dashed line), $\sigma_\Delta$ eq.(\ref{39} - dotted line) and $\sigma_{N+\Delta}$ (continuous line).}
\label{fig.8}
\end{center}
\end{figure}

The Fourier transform of the scalar form factor corresponds to a mass density, that surrounds the nucleon and
is due, in part, to the pion cloud.
As the density of the quark condensate is negative, the sign of the scalar density indicates that the pions, as Goldstone bosons,
destroy the condensate in order to exist.
In this picture, the energy density of the scalar cloud cannot exceed that of the condensate.
In fig.\ref{fig.8}, we plot the function $\sigma(r)/f_\pi^2 \mu^2$ as function of the distance, and it is possible to see that this critical
situation occurs close to 0.6 fm. 
On the other hand, the value 1/100, for which the pion field may be considere as weak, occurs around 1.5 fm, suggesting 
that this is the order of magnitude of the distances beyond which perturbative calculations may be considered as reliable.

In order to explore this picture further, we recall that the nucleon $\sigma$-term is related to $\sigma(r)$ by 
\begin{equation}
\sigma = 4\pi\;\int_0^{\infty} dr\;r^2 \sigma(r) \;.
\label{41}
\end{equation}
The evaluation of this integral, using $\sigma(r) = \sigma_N(r)+\sigma_\Delta(r)$ and replacing this function by 
$f_\pi^2 \mu^2$ whenever $\sigma(r) > 1$, yields $\sigma=46$ MeV, quite close to the value quoted in ref.\cite{GLS}.
This somewhat surprising result is associated with the factor $r^2$ in eq.(\ref{34}), that suppresses the integrand at small
distances, and seems to indicate that other mechanisms contribute little to scalar dynamics.

Another interesting aspect of expression (\ref{39}) is that it allows one to study the behaviour of the scalar form factor in the 
large $N_c$ limit, where the nucleon and the delta become both heavy and degenerate.
In that limit one has $\sigma_\Delta = 2 \sigma_N$ and hence,
\begin{equation}
\sigma(r) = \sigma_N(r) + \sigma_\Delta(r) = \frac{9\; \mu^6}{128 \pi^2}\left( \frac{g}{m} \right)^2
\left[ \mbox{\boldmath $\nabla$}_{\!\!x}\;\frac{e^{-x}}{x} \right]^2 \;.
\label{42}
\end{equation}

This expression is identical with that obtained in the framework of the Skyrme model, where 
${\cal L}_{SB}(x) = f_\pi^2 \mu^2 \cos F(x)$ and, for large distances, one has\cite{A} 
\begin{equation}
F(x) \rightarrow \left(\frac{3\; g\;\mu^2}{8\pi m f_\pi}\right) \; \frac{d}{dx} \;\frac{e^{-x}}{x}\;.
\label{43}
\end{equation}
This result, when combined with the fact that the model accounts well for the OPEP, shows that skyrmions do have 
asymptotic properties consistent with precise perturbative calculations in the large $N_c$ limit.
On the other hand, in the perturbative approach, the departure from this limit is very important, suggesting that some of the quantitative limitations of the Skyrme model may be associated with the intrinsic $N$-$\Delta$ degeneracy.

\section{Summary}

In this work we studied the central $NN$ potential at large distances and have emphasized the special role played by
the nucleon scalar form factor, which describes the mass density of its pion cloud.
The effect of this cloud over the other nucleon, represented as a multipole expansion, dominates the asymptotic interaction.
This rather simple picture accounts for about 75\% of the potential, as given by the precise phenomenological fit
by the Argonne group.

We have evaluated this leading contribution, due to the exchange of two pions, and determined its dependence
on the nucleon mass, using the relativistic formalism.
Our results coincide with those of heavy baryon chiral perturbation theory when the nucleon mass is taken to infinity, but become about 20\% smaller when the empirical mass is used.
This shows that it is important to keep the nucleon mass finite in precise descriptions of the $NN$ system.

In a recent work, Becher and Leutwyler have shown that  the scalar form factor is determined by an intermediate $\pi N$ amplitude, close to the threshold at $t=4\mu^2$, which lies almost on top of a branch point. 
As a consequence, the heavy baryon expansion does not make sense for the scalar form factor.
We  have investigated the implications of this result for the $NN$ interaction and found out that threshold 
effects can be seen at very large distances only.
Therefore, in principle, HBChPT could account numerically for the discrepancy between the heavy baryon limit and reality
found in the central potential, but this would require the extension of existing calculations to higher orders. 
This stresses the advantage of the relativistic framework, namely that it does not deal with truncated expressions and hence 
gives rise automatically to leading contributions which are already rather accurate.
A fully covariant description of the $NN$ interaction, including other degrees of freedom, will be presented elsewhere.

\vspace{5mm}
\noindent
{\large \bf Acknowledgements}\\[2mm]
We would like to thank Dr. T. Becher and Prof. H. Leutwyler for discussions about the scalar form factor and 
Dr.I.P. Cavalcante for conversations about the Skyrme model.


{}

\end{document}